\documentclass[english,twocolumn,aps,prd,amsmath,amsfonts,longbibliography,crop,tikz,superscriptaddress]{revtex4-1}
\usepackage[T1]{fontenc}
\usepackage{graphicx}
\usepackage{babel}
\usepackage{braket}
\usepackage{qcircuit}
\usepackage{xcolor}
\usepackage{lipsum}
\usepackage{blindtext}
\usepackage{physics}

\usepackage{qcircuit}
\usepackage[colorlinks,citecolor=red,urlcolor=blue,bookmarks=false,hypertexnames=true]{hyperref} 

\usepackage{fancyhdr}

\fancypagestyle{fancyplain}
\newcommand*\xbar[1]{%
  \hbox{%
    \vbox{%
      \hrule height 0.5pt 
      \kern0.3ex
      \hbox{%
        \kern-0.2em
        \ensuremath{#1}%
        \kern-0.1em
      }%
    }%
  }%
}

\date{\today}

\begin{document}

\title{Hybrid VQE-CVQE algorithm using diabatic state preparation}

\author{John P. T. Stenger}
\affiliation{U.S. Naval Research Laboratory, Washington, DC 20375, United States}
\author{C. Stephen Hellberg}
\affiliation{U.S. Naval Research Laboratory, Washington, DC 20375, United States}
\author{Daniel Gunlycke}
\affiliation{U.S. Naval Research Laboratory, Washington, DC 20375, United States}

\begin{abstract}

We propose a hybrid variational quantum algorithm that has variational parameters used by both the quantum circuit and the subsequent classical optimization.  Similar to the Variational Quantum Eigensolver (VQE), this algorithm applies a parameterized unitary operator to the qubit register.  We generate this operator using diabatic state preparation.  The quantum measurement results then inform the classical optimization procedure used by the Cascaded Variational Quantum Eigensolver (CVQE).  We demonstrate the algorithm on a system of interacting electrons and show how it can be used on long-term error-corrected as well as short-term intermediate-scale quantum computers.  Our simulations performed on IBM Brisbane produced energies well within chemical accuracy.

\end{abstract}

\maketitle
\thispagestyle{fancyplain}

\section{Introduction}


Quantum algorithms that simulate quantum systems can be designed for Noisy Intermediate-Scale Quantum (NISQ) and Fault-Tolerant Quantum (FTQ) computers.  Algorithms that require a large number of quantum logic gates must use FTQ computers.  Conversely, algorithms that require relatively few quantum logic gates, can be used on NISQ computers but are not designed for FTQ computers.  These two classes of algorithms often function very differently, which may pose a challenge when transitioning from near-term to long-term quantum computing.  

 Algorithms designed for FTQ computing include quantum phase estimation~\cite{kitaev1995} and adiabatic state preparation~\cite{Farhi2001,Smelyanskiy2002,Reichardt2004}.  Both algorithms require precise implementations of the quantum evolution operator.  Implementing such an operator requires FTQ computers, which are presently not available.  Therefore, much effort has been made to find near-term algorithms that can run on NISQ computers.  There are three groups of NISQ algorithms that optimize: (1) during the quantum computing process, including the Variation Quantum Eigensolver (VQE)~\cite{Peruzzo2014,McClean2016,OMalley2016,Kandala2017,Wang2019,McArdle2020,Arute2020,Gonthier2020,HeadMarsden2021,Cerezo2021} algorithms; (2) after the quantum computing process, including Krylov Eigensolvers~\cite{parrish2019,Stair2020,Baker2021,jamet2021,Bharti2021,Cohn2021,jamet2022,Baker2024,Yoshioka2025} and the Cascaded Variational Quantum Eigensolver (CVQE)~\cite{Gunlycke2024,Stenger2024,gunlycke2025}; and (3) both during and after, including Subspace Expansion methods~\cite{McClean2017,Colless2018,Urbanek2020,Endo2021,Motta2021,Epperly2022,Motta2024,Umeano2025}, multi-state contraction~\cite{Parrish2019b,Huggins2020}, the Variational Quantum-Neural Hybrid Eignsolver~\cite{Zhang2022} and configuration recovery~\cite{Moreno2025}. 
 
 
 We focus on VQE and CVQE.  Both VQE and CVQE utilize a variational ansatz.  Many ansatzes for both VQE and CVQE have been proposed, including the unitary coupled cluster ansatz~\cite{Peruzzo2014,Harsha2018,Kivlichan2018,Hempel2018,Romero2018,Dallaire2019,Setia2019,Lee2019,Matsuzawa2020,Sokolov2020,Yunseong2020,Bauman2020,Grimsley2020,Motta2021}, the quantum alternating operator ansatz~\cite{Farhi2014,Lloyd2018,Hadfield2019,Wang2020,Morales2020}, the hardware-efficient ansatz~\cite{Kandala2017,Barkoutsos2018,Kandala2019,Kokail2019,Ganzhorn2019,Otten2019,Gard2020,Bravo_Prieto2020,Tkachenko2021,Burton2024}, the Jastrow--Gutzwiller ansatz~\cite{Mazzola2019,Yao2021,Murta2021,Zhang_2022,Seki2022,Stenger2023}, and the Variational Hamiltonian ansatz~\cite{Wecker2015,Ho2019,Wiersema2020}.  Many of these ansatzes will struggle to scale as the problem sizes increase~\cite{Larocca2025}.  However, some of these, like the Hamiltonian ansatz, can limit onto adiabatic state preparation.   
    
In this paper, we propose a hybrid VQE-CVQE algorithm that uses the guiding-sampling ansatz~\cite{gunlycke2025}, which explicitly approximates adiabatic state preparation.  The cascaded optimization is performed by diagonalizing an effective operator in a subspace defined by the quantum measurements.  We show how this algorithm can be used during different phases of quantum computer development.  The algorithm is general enough to be useful with error-correcting quantum computers, quantum computers that have high coherence but not error corrected, near-term quantum computers, and the quantum computers available today.

\section{Method}


 \begin{figure}[t]
\vspace{2mm}
\begin{center}
\includegraphics[width=\columnwidth]{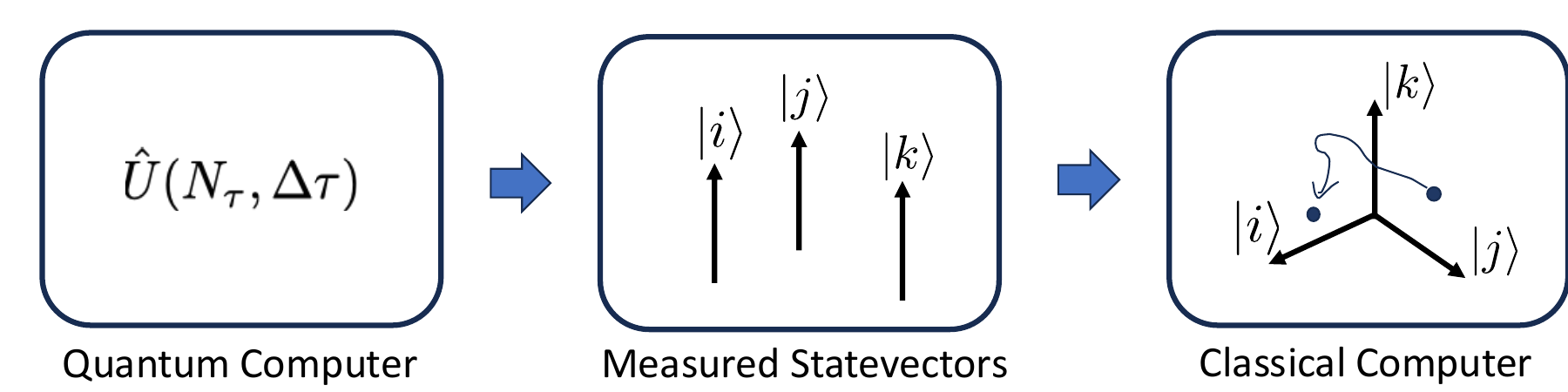}
\end{center}
\vspace{-2mm}
\caption{ The basic process for our method.  Diabatic state preparation is performed on the quantum computer to generate a guiding state.  A collection of state states are measured.  These state states form a subspace and this subspace can be optimized classically.  }  
\label{FA}
\vspace{-3mm}
\end{figure}

Our goal is to calculate the ground-state energy of a fermionic quantum system described by the Hamiltonian $\hat H$ on the Fock space $\mathcal F$.  To achieve this objective we: (1) apply a parameterized unitary operator to the quantum register to generate a guiding state $\ket{\Psi_0}$, (2) measure $\ket{\Psi_0}$ and use those measurement results to form a collection of basis states $\mathcal B$, and (3) search for the ground state of $\hat H$ on the Hilbert subspace $\mathcal{V} $ spanned by $\mathcal B$.  This method is depicted in Fig.~\ref{FA}.

\begin{table}[h]
\caption{\label{tab:algorithm} CVQE-VQE hybrid algorithm using diabatic state preparation.}
\begin{ruledtabular}
\begin{tabular}{ll}
0: Select initial values for $N_{\tau}$ and $\Delta \tau$.
\\
1: Initialize the quantum register to $\ket{\Phi_0}$. 
\\
2: Apply $\hat U(N_\tau,\Delta \tau)$ to the quantum computer.  
\\
3: Measure $\hat U(N_\tau,\Delta \tau) \ket{\Phi_0}$ and collect the set of measured 
\\
\qquad  states $\mathcal B_0$.
\\
4: Apply $\hat H$ to each state in $\mathcal B_0$,  collect the set of coupled
\\
 \qquad basis states $\mathcal B_1$, and form the union $\mathcal B = \mathcal B_0 \cup \mathcal B_1$.
\\
5: Let $\hat {H}_{\mathcal B}$ be the projection of $\hat H$ onto the subspace 
\\
\qquad $\mathcal V = \text{span}(\mathcal B)$. 
\\
6: Find the lowest eigenvalue $E_{\mathcal B}$  of $\hat {H}_{\mathcal B}$.
\\
7: If $E_{\mathcal B}$ has not converged, update $N_{\tau}$ and $\Delta \tau$ 
\\
\qquad  and return to step 1.
\\
8: Return $E_{\mathcal B}$.  
\end{tabular}
\end{ruledtabular}
\label{algorithm}
\end{table}

An algorithmic implementation of this method is listed in Table~\ref{tab:algorithm}. To construct $\ket{\Psi_0}$, we first introduce a model system described by a Hamiltonian $\hat H_0$ with a known ground state $\ket{\Phi_0}$ that we prepare on the qubit register.  Next we apply the diabatic state evolution operator
\begin{equation}
    \hat{U}(N_{\tau},\Delta \tau) = \mathcal T \prod_{\tau_i=1}^{N_{\tau}} e^{-i  \hat H(\tau_i \Delta \tau ) \Delta \tau },
    \label{UNt}
\end{equation}
with a number of time steps $N_{\tau}$, time-step duration $\Delta \tau$, a time ordering operator $\mathcal T$, and a parameterized Hamiltonian 
\begin{equation}
    \hat H(\tau) = \left(1 - \frac{\tau}{N_{\tau} \Delta \tau} \right)\hat H_0 + \frac{\tau}{N_{\tau} \Delta \tau} \hat H.
\end{equation}
This yields the guiding state $\ket{\Psi_0} = \hat U(N_\tau,\Delta \tau) \ket{\Phi_0}$.

In the limit $N_{\tau} \rightarrow \infty$ for a constant $T = N_{\tau} \Delta \tau $, the operator $ \hat{U}(N_{\tau},\Delta \tau)$ prepares the true ground state of $\hat H$. 
 Although this limit is intractable on current quantum computers, a good guiding state can still be reached using highly discretized  evolution [i.e. $\Delta \tau \gg 0$, $N_{\tau} = \mathcal {O}(1)$ ]. For more information, see Appendix~\ref{ASP} for arguments as to why highly discretized evolution could produce good guiding states and Appendix~\ref{qgtbtdaspu} for a method to decompose $U(N_{\tau},\Delta \tau)$ into quantum gates.  

 Once the guiding state is prepared, we perform a set of measurements of the guiding state on the quantum register.  The measurement outcomes $\ket{n_0}$ are basis states of the quantum register and they form the set $\mathcal B_0$.  Next we use the Hamiltonian to produce a second set $\mathcal{B}_1$ formed by all the states $\ket{n_1} $ for which there exists a state $\ket{n_0} \in \mathcal B_0$ such that $ \bra{n_1}\hat H \ket{n_0} \neq 0$. 
Lastly we combine these two sets to form the set $\mathcal B = \mathcal B_0 \cup \mathcal B_1$.
 See appendix~\ref{CS} for a comparison with the standard VQE measurement approach~\cite{Peruzzo2014}.

 The set $\mathcal{B}$ is used to form the subspace $\mathcal{V} = \text{span}(\mathcal B)$.  We expect that the ground state of $\hat H$ is well approximated within this subspace.  To generate this approximation we define the effective Hamiltonian 
 \begin{equation}
     \hat{ H}_{\mathcal B} = \sum_{\ket{n}\in \mathcal{B}}\sum_{\ket{m} \in \mathcal{B}} h_{nm} \ket{n}\!\bra{m},
     \label{barH}
 \end{equation}
  where $h_{nm} = \bra{n} \hat H \ket{m}$.  Because the dimension of $\mathcal V$ increases polynomially with systems size, we can obtain the ground-state energy $E_{\mathcal B}$ of $ \hat{H}_{\mathcal B}$ using a classical computer.

Each given pair $(N_{\tau}, \Delta \tau)$ specifies a unique guiding state $\ket{\Psi_0}$ in the CVQE guided sampling ansatz~\cite{gunlycke2025}.  This guiding state can be optimized over $(N_{\tau}, \Delta \tau)$ to improve the solution, lowering $E_{\mathcal B}$.  When $(N_{\tau}, \Delta \tau)$ is varied, our method can be understood as a hybrid VQE-CVQE method.  The optimization technique for updating $(N_{\tau}, \Delta \tau)$ can be chosen by the user.  Herein, we scan over certain cuts of the $(N_{\tau}, \Delta \tau)$ parameters space.  Note that, in our method, the variational step in CVQE is being optimized using direct diagonalization.   See appendix~\ref{OOC} for a detailed discussion on the relationship between our method and CVQE.  See appendix~\ref{aaevuas} for a discussion on how $E_{\mathcal B}$ depends on the number of shots.

\section{Demonstration}

\subsection{Model}

We demonstrate the algorithm for a spinless electronic toy model with orbitals that are equally spaced in energy and with 1-electron excitations and 2-electron interactions restricted to nearest-energy orbitals.  The Hamiltonian for this model is
\begin{equation}
\begin{split}
    \hat H = & \Delta \mu \sum_{q=0}^{Q-1} q~ \hat n_q - t \sum_{q=0}^{Q-2} \big( c^{\dagger}_q c_{q+1} + c^\dagger_{q+1} c_q \big) 
    \\
    &+ V \sum_{q=0}^{Q-2} \hat n_q \hat n_{q+1},
\end{split}
\end{equation}
where $c^{\dagger}_q$ and $c_q$ are creation and destruction operators acting on orbital q, $Q$ is the number of orbitals, $\hat n_q = c^{\dagger}_q c_q$ is the number operator, $\Delta\mu$ is the energy-level spacing, $t$ is the excitation parameter, and $V$ is the electron-electron interaction strength.  In order to build the guiding state we define the initial Hamiltonian 
\begin{equation}
    \hat H_0 = \Delta \mu \sum_{q=0}^{Q-1} q~ \hat n_q. 
\end{equation}  
The eigenstates of $\hat H_0$ can be conceptualized as electrons filling the energy levels $E_{q}^0 = q\Delta \mu$.
We initialize the quantum register to the eigenstate of $\hat H_0$ that has $N_e$ electrons filling the lowest $N_e$ energy levels.

We use the Jordan--Wigner transformation~\cite{Jordan1928} to map the electronic operators to Pauli operators~\cite{Batista2001}.  We use standard methods to generate quantum gates from the Pauli operators.  Information regarding gate generation can be found in Appendix~\ref{qgtbtdaspu}.  From these gates we obtain the quantum circuits that generate the guiding states. 

\subsection{Simulation of an 8-orbital system}
\label{so8el}


\begin{figure}[t]
\vspace{2mm}
\begin{center}
\includegraphics[width=\columnwidth]{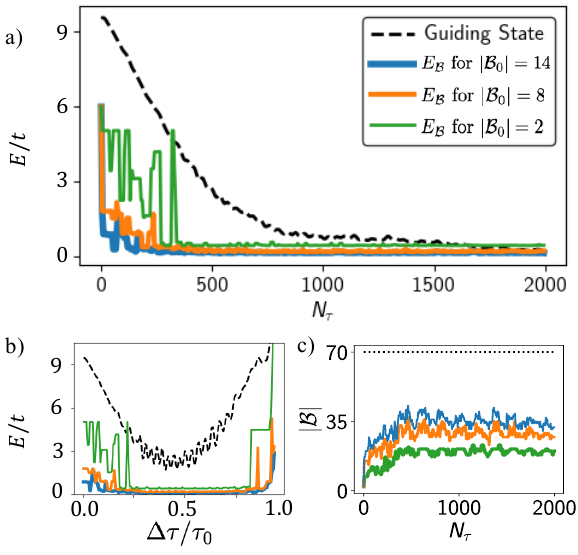}
\end{center}
\vspace{-2mm}
\caption{ Energy optimization for a $Q = 8$ orbital system with $N_e = 4$ electrons.  (a) energy as a function of the number of time steps with a duration of $\Delta\tau = 1/15~ \tau_0$ where $\tau_0 = 1/t$ is the characteristic time scale.  (b) energy as a function of the time step duration at $N_{\tau} = 100$.  (c) the total subspace after application of the Hamiltonian as a function of the number of time steps with a duration of $\Delta\tau = 1/15~ \tau_0$.  The parameters are set to $\Delta\mu = 0.75~t$, $V=t$.  While $t$ is a model parameter that sets the energy scale, a reasonable value to set our model to the molecular scale is $t = 1/15~\text{Ha}$ .  The dashed black lines represents the energy expectation value of the guiding state.  The thick blue lines represents subspace extraction using the 14 most probable states, the medium orange line represents using the 8 most probable states, and the thin green line represents using the 2 most probable states.  The dotted line in (c) marks the total size of the particle conserving state space.
}
\label{F2}
\vspace{-3mm}
\end{figure}

Figure~\ref{F2} shows results from classical simulations of a system with $Q=8$ orbitals and $N_e = 4$ electrons.  The guiding state is generated using the diabatic state preparation in Eq.~\eqref{UNt}.  The energy expectation value of the guiding state is shown as the black dashed line in Fig.~\ref{F2}a.  
The solid curves show the effective ground state energy $E_{\mathcal B}$ for various subspace dimensions.  We consider three $N_{\tau}$ regimes: a small, medium, and large regime.  In the small regime ($0 < N_{\tau} \lesssim 350 $ in Fig.~\ref{F2}a), the guiding state is far from the ground state.  However, $E_{\mathcal B}$ can still be a good approximation for certain variational parameters $(N_{\tau},\Delta \tau)$, and therefore, the hybrid VQE-CVQE algorithm can be advantageous in this regime.  In the medium regime ($350 \lesssim N_{\tau} \lesssim 2000 $  in Fig.~\ref{F2}a), the guiding state remains a relatively poor approximation of the ground state but with the energy $E_{\mathcal B}$ nearly independent of the variational parameters.  The guiding state has a  probability distribution that generates a subspace that yields a good ground-state approximation.  Thus, in this regime, CVQE can find a good approximation of the ground state without having to optimize $(N_{\tau},\Delta \tau)$.  In the large regime ($2000 \lesssim N_{\tau} $  in Fig.~\ref{F2}a), the guiding state is a good approximation of the ground state as adiabatic state preparation has been achieved.  This is indicated by the low energy of the dashed curve.  
 
Because the number of CNOT gates is proportional to the value $N_{\tau}$, these regimes determine what type of quantum computer is needed.    Because of quantum decoherence, NISQ computers are likely only going to be able to access the small and medium regimes.  By contrast, FTQ computers can access all regimes.   



 The boundaries between these regimes depend on the chosen system and the step size $\Delta \tau$.  Figure~\ref{F2}b shows the energy as a function of $\Delta \tau$ for $N_{\tau} = 100$.   Notice that the shape of the curves appear similar in Fig.~\ref{F2}a for small $N_{\tau}$ and Fig.~\ref{F2}b for small $\Delta \tau$.  That is because the unitary operator is effectively a function of the overall time $T = N_{\tau} \Delta \tau$ in the small $\Delta \tau$ limit where Trotter--Suzuki error are small.  As the optimal value of $\Delta\tau$ is both hardware and problem specific, it is useful to let $\Delta\tau$ as well as $N_{\tau}$ be variational.  

Fig.~\ref{F2}c shows the dimension of each subspace.  Because the total number of states required to cover the full 4-electron 8-orbital space is ${8 \choose 4} = 70$, the subspace dimensions are not greatly reduced.  This is typical for small quantum systems because the number of measurements needed to obtain good statistical averages produces a subspace in our method with a dimension that approaches that of the complete basis.

\subsection{Quantum computation for 50 energy levels}

\begin{figure}[t]
\vspace{2mm}
\begin{center}
\includegraphics[width=\columnwidth]{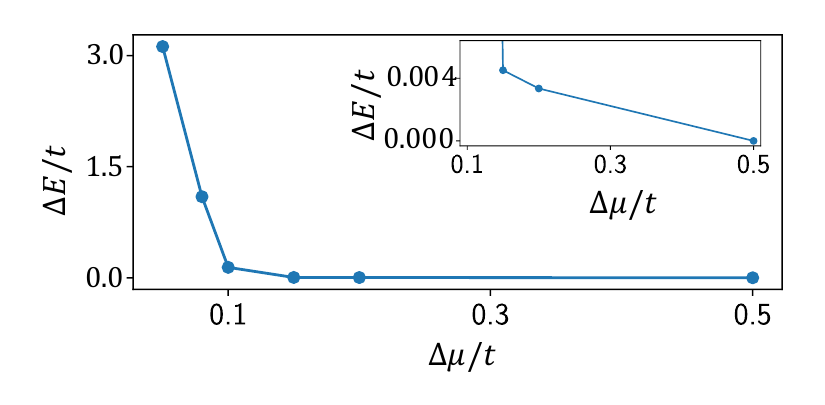}
\end{center}
\vspace{-2mm}
\caption{ The difference between the energy calculated on the quantum computer and the exact ground-state energy.  The data is represented by the circles.  The lines are drawn to guide the eye.  For this data, $Q=50$, $N_e=25$, $V=0$, $N_{\tau} = 1$, and $\Delta \tau = 1/15~\tau_0$ where $\tau_0 = 1/t$.  The inset is a magnification of the low-error data.  }  
\label{FQ}
\vspace{-3mm}
\end{figure}

 Next, consider a $Q=50$ and $N_e = 25$ system.  The size of the complete electron-preserving Fock subspace for this system is ${50 \choose 25} \approx 1.3 \times 10^{14}$.  Assuming 64-bit floating-point numbers, it requires approximately two petabytes of information to store a single state vector.  Diagonalizing a general Hamiltonian is qubic in the size of the statespace and is, thus, well beyond the reach of modern classical computers.  Instead, we run our algorithm on the IBM Brisbane quantum computer.  We set $V=0$ so that we can compare the final energy from our algorithm to known values.  The subspace dimension $|\mathcal B|$  in our method ranges from 2,000 -- 2,500, which is tiny compared to the dimension of the electron-preserving Fock subspace.  
 
 Figure~\ref{FQ} shows the energy difference $\Delta E$ between $E_{\mathcal B}$ and the ground-state energy as a function of the energy spacing $\Delta\mu$.  The error is small for all $\Delta\mu$ except for the limit $\Delta\mu \ll t$ where initializing the system to the lowest-energy state of $H_0$ is no longer justified. 
 The small error values are significant considering the error per layered gate (EPLG) for IBM Brisban is 1.9\%. 
  The error tolerance inherent in our method is due to the fact that our method is resilient to small fluctuations in the probability distribution obtained from the quantum computer.  Fluctuations in the probability distribution do not influence the final energy unless the fluctuations are large enough to add or remove a basis state to the set $\mathcal B_0$.

\begin{figure}[t]
\vspace{2mm}
\begin{center}
\includegraphics[width=\columnwidth]{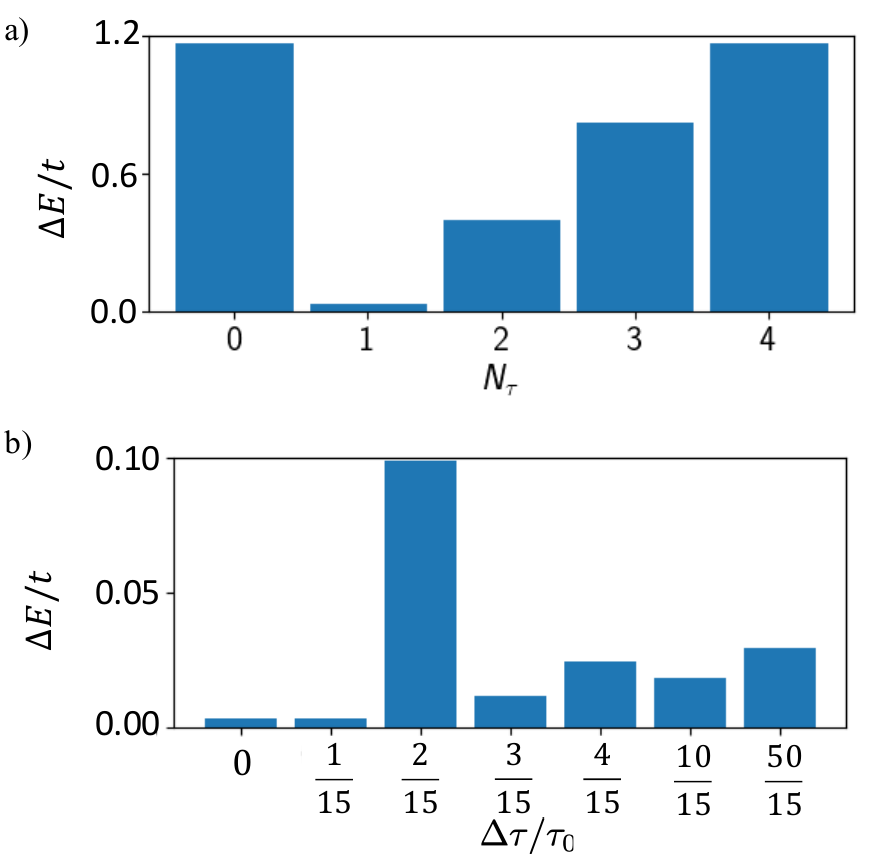}
\end{center}
\vspace{-8mm}
\caption{ Error in energy as a function of (a) the number of time steps with $\Delta \tau = 1/15~\tau_0$ where $\tau_0 = 1/t$ and (b) the duration of the time step with $N_{\tau} = 1$. In both plots, $Q=50$, $N_e=25$, $\Delta\mu = 0.2 t$, and $V=0$. 
}  
\label{F3}
\vspace{-3mm}
\end{figure}

 Figure~\ref{F3} shows the error in energy as a function of the number of time steps and the duration of the time steps.  We find the optimized solution $(N_{\tau},\Delta \tau ) = (1,1/15 \tau_0)$.  We also find that the value of $\Delta \tau$ is much less relevant than the value of $N_{\tau}$.  Notice that the scale on the y-axis of Fig.~\ref{F3}a is about an order of magnitude larger than that on Fig.~\ref{F3}b.  The hardware noise from the quantum computer plays a significant roll in these results.  Notice that at $\Delta \tau = 0$, we still find an accurate energy. From Fig.~\ref{F2}b and Ref.~\cite{gunlycke2025}, one would expect a significant increase in the error as $\Delta \tau \rightarrow 0$.  However, those are noise-free simulations.  In practice, the noise effectively applies small non-zero angles of rotation to the quantum gates even if the input instruction is zero.  
 Additionally, the fact that $N_{\tau} = 1$ is the optimal value is almost certainly due to the quantum noise at the hardware level.  The increased accuracy that one would expect by increasing $N_{\tau}$ is not observed, suggesting that the increase in error due to the larger gate count overcomes any improvement in accuracy.  
 
 In Sec.~\ref{so8el}, we described three $N_{\tau}$ regimes in which the optimal implementation of our algorithm is different. These findings using present-day quantum computers highlights a special case in which the quantum error is too high to access any $N_{\tau} > 1$.  In this case varying $(N_{\tau},\Delta \tau)$ is likely not necessary~\cite{gunlycke2025}.

 \section{Conclusion}

We have presented a hybrid VQE-CVQE algorithm that use diabatic state preparation to generate the quantum variational guiding state.  The quantum register is measured and the measurements are processed to form a collection of basis states, which are then used to form a subspace on which we represent the system Hamiltonian.  The lowest eigenvalue of this Hamiltonian is then calculated using a classical computer.  

We find that this hybrid algorithm performs better for NISQ computers than either VQE or CVQE on their own.  We have described three regimes for which the optimal implementation of the algorithm changes.  Although we demonstrate the presence of these regimes only for our toy model, we expect that these regimes are present for many physically relevant models.  For the quantum computers that are currently available, minimizing the circuit is the most critical consideration, thus $N_{\tau} = 1$ gives the best results for large systems.  In this case, there is no need to update the variational parameters.

  In the near term,  allowing the variational parameters to update could help optimize the energy.  In the medium term, the energy can optimized using CVQE without changing the variational parameters.  Eventually, when we have FTQ computers, we can do adiabatic state preparation directly without the need for any additional optimization.  Our generalized method is effective in all of these regimes.  

 \section{Acknowledgement}

This work has been supported by the Office of Naval Research (ONR) through the U.S. Naval Research Laboratory (NRL).  We acknowledge QC resources from IBM through a collaboration with the Air Force Research Laboratory (AFRL).

\appendix

 \section{Diabatic State Preparation}
 \label{ASP}

We need a $\hat{U}$ that evolves the state of the quantum register towards the ground state of $\hat H$ so that $\mathcal B$ contains the necessary states to build a good approximation of the ground state.  If we had access to an error-free quantum computer, we could perform adiabatic state preparation
\begin{equation}
    \hat U_A(\tau) = \mathcal T e^{-i\int_{0}^{\tau} \text{d}\tau_1 \hat H(\tau_1)} .
\end{equation}
where 
\begin{equation}
    \hat H(\tau) = \left(1 - \frac{\tau}{T} \right) \hat H_0 + \frac{\tau}{T} \hat H ,
\end{equation}
where  $H_0$ is trivial to diagonalize and $T >> 1/|\hat H|$ is a long time range.

To apply this operator on a quantum computer we have to discretize time as in Eq.~\eqref{UNt} of the main text where $T = \Delta \tau N_{\tau}$.  In the limit $N_{\tau} \rightarrow  \infty$ and $\tau = T$  we recapture the evolution operator $\hat{U}(N_{\tau} \rightarrow \infty,\Delta \tau = T/N_{\tau}) = \hat{U}_A(T)$. By initializing the quantum computer to an eigenstate of $\hat H_0$ and applying $\hat U_A(T)$, we are guaranteed to prepare the exact ground state of $\hat H$ when $T \rightarrow \infty$, assuming the spectrum of $\hat H(\tau)$ is gapped throughout.  However, fast error rates prevent us from approaching the $N_{\tau} \rightarrow \infty$ limit. 

 \begin{figure}[t]
\vspace{-2mm}
\begin{center}
\includegraphics[width=\columnwidth]{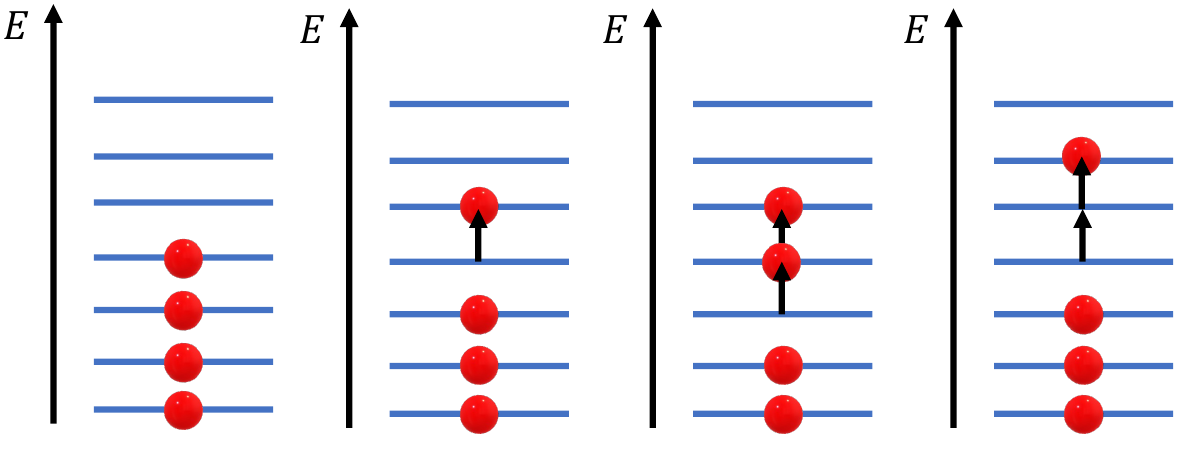}
\end{center}
\vspace{-2mm}
\caption{  Eigenstates for an example $H_0$. 
 The spheres represent electrons.  The lines represent energy levels.  On the left, the lowest energy levels are filled with electrons.  In the second from left, one excitation has occurred.  The two right most graphs show two excitations.  }  
\label{FAp}
\vspace{-3mm}
\end{figure}

Fortunately, for our purposes, we only need to reach an approximation of the correct subspace, not of the correct eigenstate itself. Figure~\ref{FAp} shows energy levels filled with four electrons in different excitation configurations.  We expect that the ground state should be a linear combination of the lower-order excitations. These lowest order excitation are well captured by diabatic state preparation.  Let us compare a single time step $N_{\tau} = 1$ to the full adiabatic evolution
\begin{equation}
\begin{split}
    \hat U(1,T) = e^{-i \hat H_0 T} + \sum_{m=1}^{\infty}\sum_{n=1}^{\infty} (-iT)^n \Pi_{n-m,m}(\hat H_0,\hat H_1),
    \\
    \hat U_A(T) = e^{-i \hat H_0 T} + \sum_{m=1}^{\infty}\sum_{n=1}^{\infty} (-iT)^n \tilde \Pi_{n-m,m}(\hat H_0,\hat H_1), 
\end{split}
\end{equation}
where $\hat H_1 = \hat H - \hat H_0$ is the non-trivial part of $\hat H$ and both $\Pi_{n,m}(\hat H_0,\hat H_1)$ and $\tilde \Pi_{n,m}(\hat H_0,\hat H_1)$ are sums over all permutations of operator strings with a number of $\hat H_0$ operators equal to $n$ and a number of $\hat H_1$ operators equal to $m$ but with different weights. For $\Pi_{n,m}(\hat H_0,\hat H_1)$, all operators have the weight $w = 1/(n+m)!$.  For example
\begin{equation}
    \Pi_{2,1}(\hat H_0,\hat H_1) = \frac{1}{3!} (\hat H_0 \hat H_0  \hat H_1 + \hat H_0 \hat H_1 \hat H_0   + \hat H_1 \hat H_0 \hat H_0   ).
\end{equation}
For $\tilde \Pi_{n,m}(\hat H_0,\hat H_1)$, the weights depend on the position of the $\hat H_1$ operators in the operator string.  In general, the weight is
\begin{equation}
    \bar w = \frac{1}{\gamma_1(\gamma_1 + \gamma_2)(\gamma_1 + \gamma_2 + \gamma_3)\ldots},
    \label{barw}
\end{equation}
where $\gamma_j =2$ if the operator $j$ positions from the left is $\hat H_1$ and $\gamma_j =1$ otherwise (a derivation for this expression is found below in \ref{witeo}).  For example,
\begin{equation}
    \tilde \Pi_{2,1}(\hat H_0,\hat H_1) =  \frac{1}{24}\hat H_0 \hat H_0  \hat H_1 + \frac{1}{12}\hat H_0 \hat H_1 \hat H_0   + \frac{1}{8}\hat H_1 \hat H_0 \hat H_0   .
\end{equation}
We see that both $\hat U(1,T)$ and $\hat{U}_A(T)$ contain the exact same operators with the general rule that the longer the operator string, the smaller the weight.  The difference is that $\hat U_A(T)$ turns on $\hat H_1$ slowly and so the operator strings with $\hat H_1$ towards the beginning (towards the right) are less relevant.  

We are only interested in the subspace that contains the ground state and not on the specific weights of each state.  Therefore, we expect that $\hat U(1,T)$ is often sufficient for our purposes as it produces the same excitations as $\hat U_A(T)$ with a similar ordering of the wieghts.

\subsection{Weights in the evolution operator}
\label{witeo}
The full-adiabatic evolution operator can be written as a sum over integrals
\begin{equation}
    \hat U_A(T) = \sum_{n=0}^{\infty}(-i)^n \hat{I}_n(T),
\end{equation}
where each integral is 
\begin{equation}
    \hat{I}_n(\tau) = \int_0^\tau \text{d}\tau_1 \hat{H}(\tau_1)I_{n-1}(\tau_1),
\end{equation}
with $\hat{I}_0(\tau) = \boldsymbol{1}$.  Let us perform the first few integrals and then the pattern for the weights becomes clear
\begin{equation}
    \begin{split}
        \hat I_1(\tau) =& \frac{1}{1}\hat H_0 \tau + \frac{1}{2} \hat H_1 \frac{\tau^2}{T} ,
        \\
        \hat I_2(\tau) =& \frac{1}{2\cdot 1} \hat H_0 \hat H_0 \tau^2 + \frac{1}{3\cdot 2}  \hat H_0 \hat H_1 \frac{\tau^3}{T} ,
        \\
        +& \frac{1}{3\cdot 1} \hat H_1 \hat H_0 \frac{\tau^3}{T} + \frac{1}{4\cdot 2} \hat H_1 \hat H_1 \frac{\tau^4}{T^2},
        \\
        \hat I_3(\tau) =&  \frac{1}{3\cdot 2\cdot 1} \hat H_0 \hat H_0 \hat H_0 \tau^3 + \frac{1}{4\cdot 3\cdot 2} \hat H_0 \hat H_0 \hat H_1 \frac{\tau^4}{T},
        \\
         +& \frac{1}{4\cdot 3\cdot 1} \hat H_0 \hat H_1 \hat H_0 \frac{\tau^4}{T} + \frac{1}{5\cdot 4\cdot 2} \hat H_0 \hat H_1 \hat H_1 \frac{\tau^5}{T^2},
        \\
         +& \frac{1}{4\cdot 2\cdot 1} \hat H_1 \hat H_0 \hat H_0 \frac{\tau^4}{T} + \frac{1}{5\cdot 3\cdot 2} \hat H_1 \hat H_0 \hat H_1 \frac{\tau^5}{T^2},
         \\
         +& \frac{1}{5\cdot 3\cdot 1} \hat H_1 \hat H_1 \hat H_0 \frac{\tau^5}{T^2} + \frac{1}{6\cdot 4\cdot 2} \hat H_1 \hat H_1 \hat H_1 \frac{\tau^6}{T^3}.
    \end{split}
\end{equation}

We see that the weights are coming from integration over $\tau$.  Each integration increases the power of $\tau$ by one but also each $\hat H_1$ increases the power of $\tau$ by one.  Therefore, we arrive at the expression for the weight in Eq.~\eqref{barw}.

\section{Quantum Gates To Build The Diabatic State Preparation Unitary}
\label{qgtbtdaspu}

\begin{figure}[h]
\vspace{2mm}
\begin{center}
\includegraphics[width=\columnwidth]{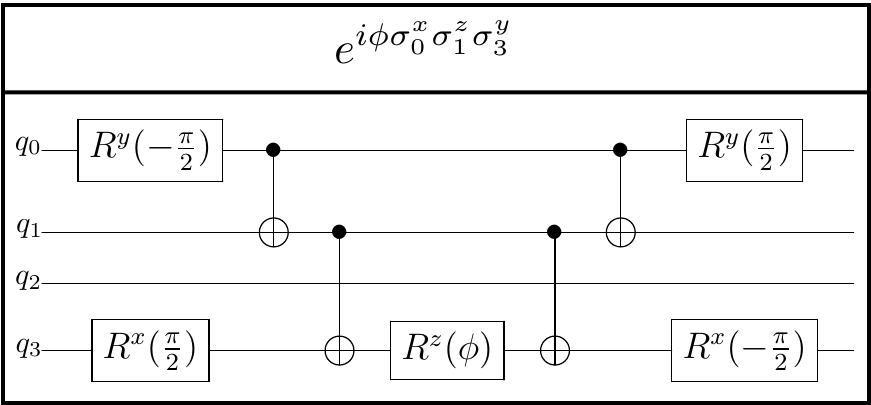}
\end{center}
\vspace{-2mm}
\caption{ Example quantum circuit to implement the unitary operator $e^{i\phi\sigma^x_0 \sigma^z_1 \sigma^y_3}$. }  
\label{FAp2}
\vspace{-3mm}
\end{figure}

The unitary operator $\hat U(N_{\tau},\Delta \tau)$ can be decomposed into quantum gates.  First, we use the JW transformation and the Trotter--Suzuki decomposition to write the operator as a product of exponentials of Pauli strings.  Let $\sigma^a_q$ be the Pauli-a operator acting on qubit $q$ and let $R^a_q(\phi)$ be the rotation around the a-axis of the block sphere for qubit $q$.  A general exponential of Pauli strings can be decomposed into quantum gates in the following way:  
\begin{itemize}
    \item On the last qubit with a non-identity Pauli string, insert an $R^z_q(\phi)$ gate with an angle equal to the weight of the Pauli string.
    \item Sandwich the $R^z_q(\phi)$ gate between two chains of CNOT gates that connect all of the qubits with non-identity Pauli operators.  
    \item For each qubit with a $\sigma^x_q$ Pauli operator, sandwich the circuit with an $R^y_q(-\pi/2)$ to the left and an $R^y_q(\pi/2)$ to the right.
    \item For each qubit with a $\sigma^y_q$ Pauli operator, sandwich the circuit with an $R^x_q(\pi/2)$ to the left and an $R^x_q(-\pi/2)$ to the right.
\end{itemize}
Figure~\ref{FAp2} shows an example of a circuit to create the exponential of the Pauli string $\sigma^x_0 \sigma^z_1 \sigma^y_3$.

\section{Collecting Shots}
\label{CS}

We describe two methods for collecting the shots from the quantum computer and, subsequently, using those shots to create a subspace.  In the first method we divide the Hamiltonian into a sum of Pauli strings and then
collect shots in the diagonalized basis of each Pauli string.  This first method is the standard for calculating the expectation values of Hamiltonians on a quantum computer.  In the second method, we collect the shots only in the original basis of the quantum computer.  We use the second method in the main text of the manuscript.  We show that the two methods are equivalent for k-local Hamiltonians.  

For the first method, we partition the Hamiltonian into a sum of $L$ Pauli strings
\begin{equation}
    \hat H = \sum_{l=1}^L \hat P_l,
    \label{Pl}
\end{equation}
where each $\hat P_l$ is a weighted Pauli string.  Each $\hat P_l$ has an associated rotation operator $\hat R_l$, which diagonalizes $\hat P_l$
\begin{equation}
    \hat R_l \hat P_l \hat R^{\dagger}_l = \sum_n E_{nl}\ketbra{n}{n},
\end{equation}
where $E_{nl}$ are the eigenvalues of $\hat P_l$. 

Let the quantum computer be in a state $\ket{\Psi_0}$. In this first method, we apply each rotation to the quantum state before measurement.  After a rotation $\hat R_l$ the amplitude of the basis state $\ket{n}$ is
\begin{equation}
    \tilde{A}_{nl} = \bra{n}\hat R_l\ket{\Psi_0}.
\end{equation}
When necessary for clarity, we refer to $\{n_l\}=\{\hat R_l \ket{n}\}$ as a rotated basis set and $\{\ket{n}\}$ as the unrotated basis.  
Let us imagine that we find a shot in state $\ket{n_l}$ if $|\tilde A_{nl}| > \epsilon$ for some cutoff threshold $\epsilon$.  Then the set of states that we obtain from the quantum computer are
\begin{equation}
   \tilde{\mathcal{B}}_{0l} = \{ \ket{n_l} :  |\tilde{A}_{nl}|>\epsilon \}.
\end{equation}
However, we need to form a single basis using all of the collected measurements.    Thus, we form the set of all unrotated basis states that might have contributed to each measured rotated basis state
\begin{equation}
    \tilde{\mathcal{B}}_l = \{ \ket{m} : r^{l}_{mn} \neq 0 \text{ and }  |\tilde{A}_{nl}|>\epsilon \},
\end{equation}
where $r^l_{mn} = \bra{m}\hat R^{\dagger}_l\ket{n_l}$.  The total set of states is the union of all $\tilde{\mathcal{B}}_l$
\begin{equation}
    \tilde{\mathcal{B}} = \bigcup_l \tilde{\mathcal{B}}_l.
\end{equation}
Note that if we are simply calculating expectation values as in VQE then we do not have to form $\tilde{ \mathcal B}$.  However, $\tilde{\mathcal B}$ is necessary to perform the CVQE step.    This completes our description of the first method.

In the second method we do not perform any rotations before the measurement.  Therefore, the amplitude of the basis state $\ket{n}$ is simply
\begin{equation}
   \breve{A}_n = \braket{n}{\Psi}.
\end{equation}
We form the initial set of states from this amplitude
\begin{equation}
    \mathcal{B}_0 = \{ \ket{n}: | \breve{A}_n| > \epsilon \},
\end{equation}
where we have again assumed that we collect shots in all states with amplitudes bigger than $\epsilon$.  Using this initial set alone, we are likely to miss important properties of the Hamiltonian.  Therefore, we find the full set of states by hitting the initial set with the Hamiltonian 
\begin{equation}
    \mathcal{B} = \{ \ket{m}: h_{mn}\neq 0 \text{ and } | \breve{A}_n| > \epsilon \},
    \label{EC9}
\end{equation}
where $h_{nm} = \bra{n}\hat H\ket{m}$ and $\hat H$ is defined such that $h_{nn} \neq 0$ for all $n$.  This completes our description of the second method.  Next, we compare the two methods.  

It is difficult to compare the two sets $\mathcal{B}$ and $\tilde{\mathcal{B}}$ as they are written.  To help make the comparison, let us define the total amplitudes $\bar A_n$ and $A_n$ such that if $\bar A_n \neq 0$ then $\ket{n}\in \tilde{\mathcal{B}}$ and if $A_n\neq 0$ then $\ket{n}\in \mathcal{B}$.  
For the first method we have
\begin{equation}
    \bar A_n = \sum_l \sum_m r^l_{mn}(\tilde A_{ml})|_{>\epsilon},
\end{equation}
  for the second method we have
\begin{equation}
    A_n = \sum_{m} h_{nm} ( \breve{A}_m)|_{>\epsilon},  
\end{equation}
where we use the notation
\begin{equation}
    (B)|_{>\epsilon} = 
    \begin{cases}
        B & \text{if } B > \epsilon \\
        0 & \text{if } B < \epsilon 
    \end{cases}.
\end{equation}
We can better compare these two expressions if we use the identities
\begin{equation}
    \tilde A_{al} = \sum_m (r^l_{ma})^* \braket{m}{\Psi},
\end{equation}
and 
\begin{equation}
    h_{nm} = \sum_l \sum_a r^l_{na}(r^l_{ma})^* E_{al}.
\end{equation}
Plugging in these two identities, we have 
\begin{equation}
    \bar A_n = \sum_l \sum_a  r^l_{na} \left[ \sum_m (r^l_{ma})^* \braket{m}{\Psi} \right]\Bigg|_{>\epsilon},
\end{equation}
and 
\begin{equation}
    A_n = \sum_l  \sum_a  E_{al} r^l_{na}  \sum_m   (r^l_{ma})^* \big[\braket{m}{\Psi} \big]|_{>\epsilon} .
\end{equation}
The value of $E_{al}$ is insignificant as we are only concerned with whether or not the probabilities are non-zero.  
The important difference is the evaluation of the cutoff.  In the first method, the cutoff is evaluated after the summation over $m$.  In the second method, the cutoff is evaluated before the summation over $m$.  The effect of the cutoff, therefore, depends on how many non-zero values $r^l_{ma}$ there are for a given $l$ and $a$.  For a Pauli string that contains $\chi_l$ Pauli-X and Pauli-Y operators, there are $N_r = 2^{\chi_l}$ non-zero $r^l_{ma}$ terms.


 We can think of the second method as effectively increasing the cutoff (i.e. $\epsilon \rightarrow N_r \epsilon$).  This can be compensated for by increasing the number of shots.  If we take $N_s$ times the number of measurements then $\epsilon \rightarrow \epsilon/ N_s$.  For typical fermionic Hamiltonians used in quantum chemistry $\chi_l \leq 4$.  Thus, $N_r \leq 16$.  Therefore, we need only take sixteen times the number of shots for the second method to achieve the same accuracy as the first method.  Furthermore, the first method requires $L$ times as many circuits to achieve this increased accuracy.  Even for simple molecules, we can have $L\approx 1000$.  So it is often much more efficient to use the second method.  We use the second method throughout the main text.

\section{Overview of CVQE}
\label{OOC}


 In the CVQE algorithm, one uses an ansatz of the form 
\begin{equation}
    \ket{\Psi(\theta)} = e^{i\hat \lambda(\theta)}\hat{U}\ket{\Phi_0},
\end{equation}
where $\hat \lambda$ is an operator, $\theta$ is a collection of parameters, $\hat{U}$ is a unitary operator, and $\ket{\Phi_0}$ is some initialization state.  The state 
\begin{equation}
    \ket{\Psi_0} = \hat{U} \ket{\Phi_0},
\end{equation}
is simulated on a quantum computer.  We calculate the energy expectation value 
\begin{equation}
    E(\theta) = \frac{\Upsilon(\theta)}{\Lambda(\theta)},
\end{equation}
where
\begin{equation}
\begin{split}
    &\Upsilon(\theta) = \bra{\Psi_0}e^{-i\lambda^*(\theta)}\hat H e^{i\lambda(\theta)}\ket{\Psi_0}, \\
    &\Lambda(\theta) = \bra{\Psi_0}e^{-i\lambda^*(\theta)}e^{i\lambda(\theta)}\ket{\Psi_0},
\end{split}
\end{equation}
in terms of the measured probability distributions 
\begin{equation}
\label{probability}
    |\tilde A_{nl}|^2 = |\bra{\Psi} \hat R^{\dagger}_l\ket{n}|^2,
\end{equation}
for various rotations $\hat R_l$.  An important insight of CVQE is that the number of required $\hat R_l$ to calculate $\Upsilon(\theta)$ and $\Lambda(\theta)$ is the same as the number required to measure an expectation value of $\hat H$, as long as $\hat \lambda(\theta)$ is diagonal in the unrotated basis $\{\ket{n}\}$.

\subsection{Comparison between CVQE and our method}

The state created on the quantum computer in CVQE $\ket{\Psi_0}$ acts as our guiding state for particular values of $N_{\tau}$ and $\Delta \tau$.  Instead of calculating $\Upsilon$ and $\Lambda$ from the probability distributions \eqref{probability}, we project the Hamiltonian onto a subspace.  However, the final ansatz is the same in both CVQE and our method.  Let $u_{n0} = \bra{n} \hat U \ket{0}$, then the guiding state is 
\begin{equation}
    \ket{\Psi_0} = \sum_{n\in \mathcal N} u_{n0} \ket{n}.
\end{equation}
Applying the second piece of the CVQE ansatz gives
\begin{equation}
    \ket{\Psi(\theta)} = \sum_{n\in \mathcal N} e^{i\hat\lambda(\theta)} u_{n0} \ket{n}.
\end{equation}
If we choose $\hat \lambda(\theta)$ such that $\hat \lambda(\theta)$ is diagonal and $\bra{n}e^{i\hat \lambda(\theta)}\ket{n} = 0$ if $\ket{n} \notin \mathcal{B}$, from Eq.~\eqref{EC9}, and redefine our variational parameter as $\phi_n = \bra{n}e^{i\hat \lambda(\theta)}\ket{n}u_{n0}$ then the variational state is 
\begin{equation}
    \ket{\Psi_{\mathcal B}(\phi)} = \sum_{n\in B} \phi_{n} \ket{n},
\end{equation}
which covers the state subspace $\mathcal V$.  One could variationally optimize $\phi$, however, we know that the optimal solution corresponds to the solution found by diagonalizing the effective Hamiltonian.  We consider diagonalization to be a method for optimizing the variational problem.  Thus, for a set values ($N_{\tau}$,$\Delta \tau$), our method is CVQE with the specific ansatz.  However, if either $N_{\tau}$ or $\tau$ are allowed to vary then our method can be thought of as a hybrid VQE-CVQE method.  


\subsection{Effective Hamiltonian}

We can optimize $\ket{\Psi_{\mathcal B}(\phi)}$ variationally, however, we can find the exact minimum from diagonalizing an effective Hamiltonian instead.  Let us define an effective Hamiltonian on the subspace as in Eq.~\eqref{barH} of the main text.
As the size of the basis $\mathcal{B}$ is subexponential with the number of qubits, we can use a classical computer to diagonalize the effective Hamiltonian 
\begin{equation}
     \hat{H}_{\mathcal B} \ket{E_{\mathcal B n}} = E_{\mathcal B n}\ket{E_{\mathcal B n}},
\end{equation}
where $\ket{E_{\mathcal B n}}$ are the eigenstates and $E_{\mathcal B n}$ are the eigenvalues of $ \hat{H}_{\mathcal B}$.  

We can show that the ground state $E_{\mathcal B} \equiv E_{\mathcal B 0}$ is the minimum possible expectation value of $\hat H$ with $\ket{\Psi_{\mathcal B}(\phi)}$.  We know that $\ket{E_{\mathcal B n}}$ forms a complete basis of the subspace $\mathcal V = \text{span}(\mathcal B)$, therefore, we can write the state as
\begin{equation}
    \ket{\Psi_{\mathcal B}(\phi)} = \sum_n \psi_{\mathcal B n}(\phi) \ket{E_{\mathcal B n}},
\end{equation}
where $\psi_{\mathcal B n}(\phi) = \bra{E_{\mathcal B n}} \ket{\Psi(\phi)}$ are complex coefficients.  Thus, the variational principle applies
\begin{equation}
    \bra{\Psi_{\mathcal B}(\phi)}  \hat{H}_{\mathcal B} \ket{  \Psi_{\mathcal B}(\phi)} = \sum_n |\psi_{\mathcal B n} (\phi)|^2 E_{\mathcal B n} \geq  E_{\mathcal B 0} .
\end{equation}
Furthermore, the expectation value of $\hat H$ with $\ket{ \Psi_{\mathcal B}(\phi)}$ is the same as that of $ \hat{H}_{\mathcal B}$
\begin{equation}
    \begin{split}
        E_{\mathcal B 0} \leq& \bra{\Psi_{\mathcal B}(\phi) }  \hat{H}_{\mathcal B} \ket{ \Psi_{\mathcal B}(\phi)} 
        \\
        = & \sum_{nm\in B} \phi^{*}_{n} \phi_{m} \bra{n} \Big( \sum_{pq\in B} \bra{p} \hat{H} \ket{q} \ket{p}\!\bra{q} \Big) \ket{m}
        \\
        = & \sum_{nm\in B} \phi^{*}_{n} \phi_{m}\bra{n} \hat{H} \ket{m} 
        \\
        = & \bra{\Psi_{\mathcal B}(\phi) } \hat{H} \ket{\Psi_{\mathcal B}(\phi)} .
    \end{split}
\end{equation}

\section{Approximating an expectation value using a subspace}
\label{aaevuas}

\begin{figure}[h]
\vspace{2mm}
\begin{center}
\includegraphics[width=\columnwidth]{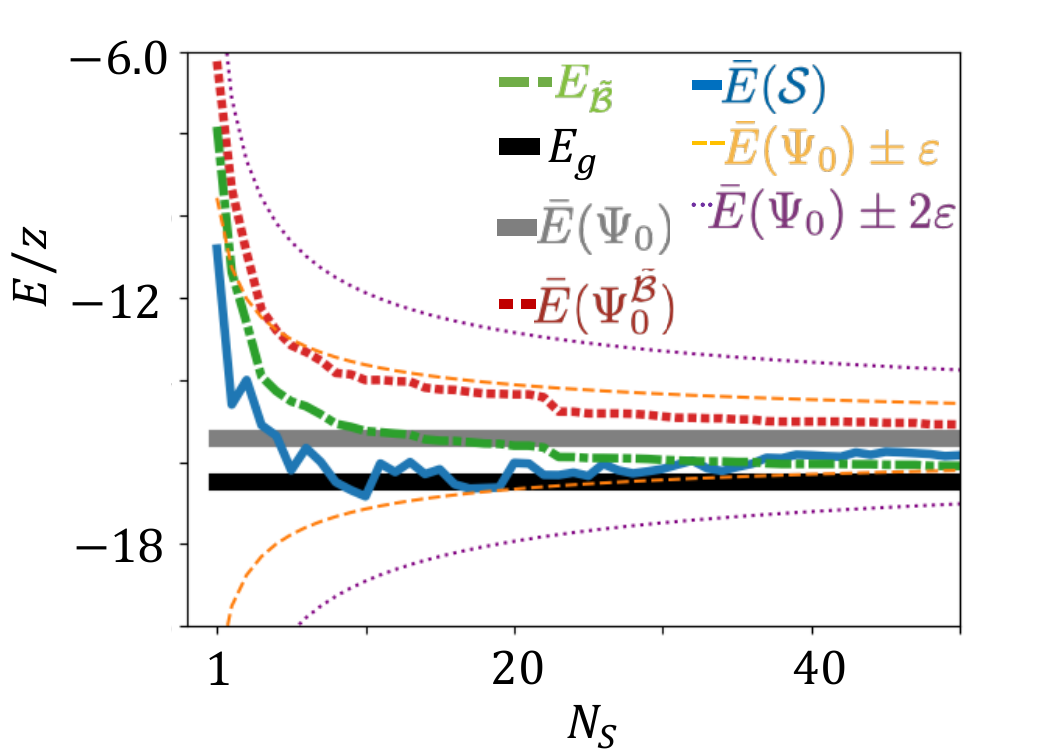}
\end{center}
\vspace{-2mm}
\caption{ Energy expectation values as a function of the number of shots.  The models used is $\hat H = \sum_{q=0}^7 (z \hat\sigma^z_q + x \hat\sigma^x_q)$ for $x = 1.8z$.  The dash-dot green curve $E_{\tilde{\mathcal B}}$ is the CVQE energy, the thick black line $E_g$ is the ground-state energy, the thick gray line $\bar E(\Psi_0)$ is the guiding state energy, the dashed red line $\bar E(\Psi_0^{\tilde{\mathcal B}})$ is the energy of the guiding state projected onto the subspace $\tilde{\mathcal B}$, the blue curve $\bar E(\mathcal S)$ is the sample distribution, and the thin dashed yellow curves and the thin dotted purple curves are defined by $\epsilon$, which is the standard error.}  
\label{FA_R}
\vspace{-3mm}
\end{figure}

The effective ground-state energy $E_{\tilde{\mathcal{B}}}$ is bounded from above based on the number of shots $N_{S}$. To show this, we use several different energy expectation values.  The energies involved in this discussion are plotted against $N_S$ in Fig.~\ref{FA_R}, for a simple model Hamiltonian. The overall strategy is to find a state vector that can be built from $\tilde{\mathcal B}$ and that is within statistical accuracy of the guiding-state energy.  We show that such a state vector is found by projecting the guiding state to the basis $\tilde{\mathcal B}$.

Let us define a general energy expectation value 
\begin{equation}
    \bar E(\mathcal D) = \sum_l \sum_n d_{nl} E_{nl},
\end{equation}
given a distribution $\mathcal D$ with elements $d_{nl}$ where $l$ indexes the Pauli-string basis and $n$ indexes the basis states.  We also need to use the conditional expectation value.  Given a condition $\mathcal C$ on the distribution, we can define the conditional expectation value
\begin{equation}
    \bar E(\mathcal D | \mathcal C)  = \sum_l \frac{1}{\text{norm}_l(\mathcal D|\mathcal C)}\sum_{n\in C_l} d_{nl} E_{nl},
\end{equation}
where $\text{norm}_l(\mathcal D|\mathcal C) = \sum_{n\in C_l} d_{nl}$ is the normalization and $C_l$ are the allowed indices in the $l$ bases given the condition $\mathcal C$.  To be precise, $n_l \in C_l$ iff there is an $n$ that satisfies $\mathcal C$ and $\bra{n_l}\hat R_l \ket{n} \neq 0$.  While each $C_l$ may be different for each $l$ based on $\hat R_l$, they are all generated from the same set of basis states.  Thus, $\bar E(\mathcal D|\mathcal C)$ can be understood as the energy expectation value from the projection of the distribution $\mathcal D$ to the basis states that satisfy $\mathcal C$. 

There are two relevant distributions for our purposes.  One is the guiding state distribution $\Psi_0$ with elements $|\braket{n_l}{\Psi_0}|^2$ and the other is the sample distribution $\mathcal S$ whose elements $N_{nl}/N_S$ are the number of shots $N_{nl}$ that returned $\ket{n_l}$ from sampling $\Psi_0$ on the quantum computer divided by the total number of shots $N_S$.   From basic statistics~\cite{Wecker2015b,Rubin2018} we know that the expectation value from the sample distribution is within the standard error of the expectation value of the guiding state distribution
\begin{equation}
    |\bar E(\mathcal S) - \bar E(\Psi_0)| \lesssim \varepsilon ,
    \label{E3}
\end{equation}
where $\varepsilon = \sqrt{\sum_l E_{0l}^2/N_S}$ is the standard error.  We also know that for a general condition $\mathcal C$, the conditional distributions are within the conditional standard error
\begin{equation}
    |\bar{E}(\mathcal S|\mathcal C) - \bar{E}(\Psi_0|\mathcal C)| \lesssim \varepsilon_{\mathcal C},
    \label{E4}
\end{equation}
where $\varepsilon_{\mathcal C} = \sqrt{\sum_l E_{0l}^2/N_S^{\mathcal C}}$ is the conditional standard error and $N_S^{\mathcal C}$ is the number of shots that satisfy $\mathcal C$.  

The condition that will allow us to bound $E_{\tilde{\mathcal{B}}}$ is that the basis states fall within $\tilde{\mathcal B}$. Let us refer to this condition as the $\tilde{\mathcal B}$ condition.  Because we use the sample distribution to build $\tilde{\mathcal B}$, all the shots satisfy the condition $N_S^{\tilde{\mathcal B}} = N_S$ and, thus, $\bar E(S|\tilde{\mathcal B}) = \bar E(S)$.  Furthermore, $\bar E(\Psi_0|\tilde{\mathcal B}) = \bar E(\Psi_0^{\tilde{\mathcal B}})$ where $\Psi_0^{\tilde{\mathcal B}}$ is the  distribution of $\ket{\Psi_0^{\tilde{\mathcal B}}}$ the guiding state projected onto the basis $\tilde{\mathcal{B}}$ and normalized.  Using these facts, Eq.~\eqref{E3}, and Eq.~\eqref{E4}, with the condition $\tilde{\mathcal B}$, we see that $\bar E(\mathcal S)$ is within the standard error of both $\bar E(\Psi_0)$ and $\bar E(\Psi_0^{\mathcal B})$.  Therefore, there is a bound between the guiding-state energy and the projected-guiding-state energy
\begin{equation}
    |\bar{E}(\Psi_0^{\tilde{\mathcal B}}) - \bar{E}(\Psi_0)| \lesssim 2 \varepsilon.
    \label{E5}
\end{equation}
Because $\ket{\Psi_0^{\tilde{\mathcal B}}}$ is confined to $\tilde{\mathcal B}$, the variational principle demands that $\bar{E}(\Psi_0^{\tilde{\mathcal B}}) \geq E_{\tilde{\mathcal{B}}}$. Thus, using Eq.~\eqref{E5} and the variational principle, we have the bound
\begin{equation}
    \bar{E}(\Psi_0) + 2\varepsilon \gtrsim E_{\tilde{\mathcal B}} \geq E_g,
\end{equation}
where $E_g$ is the ground-state energy of $\hat H$.  Therefore, the CVQE energy converges somewhere between the guiding-state energy and the ground-state energy at a rate of $1/\sqrt{N_S}$ at worst.

\section{Device Specification}

In Tables~\ref{tab:brisbane1},~\ref{tab:brisbane2},~and~\ref{tab:brisbane3} we show the device specifications for IBMQ Brisbane on which the quantum computation was performed.  Out of the 125 qubits, we use the first 50.  The native gates are Pauli-Z rotations, $\sqrt{X}$ gates, and Echoed Cross Resonance (ECR) gates.  In the tables, the ECR gates are labeled by an arrow.  The arrow indicates the orientation of the neighboring target qubit.  The arrow $\leftarrow$ ($\rightarrow$) indicate that the target qubit is one index lower (higher) than the control qubit.  The arrow $\uparrow$ ($\downarrow$) indicate that the target qubit is directly above (below) the control qubit.  

\begin{table*}[ht]
\caption{\label{tab:brisbane1} Specifications of IBMQ Brisbane. }
\begin{ruledtabular}
\begin{tabular}{lccccccccc}
qubit & Frequency (GHz) & T1 ($\mu$s) & T2 ($\mu$s) & Readout Error (\%) & $\sqrt(X)$ (\%) & ECR $\uparrow$ (\%) & ECR $\downarrow$ (\%) & ECR $\leftarrow$ (\%) & ECR $\rightarrow (\%)$  \\  $q_{0}$ & 4.72 & 291.7 & 80.1 & 2.37 & 0.01 &   &   &   &   \\  $q_{1}$ & 4.82 & 364.7 & 289.5 & 2.25 & 0.01 &   &   & 0.44 &   \\  $q_{2}$ & 4.61 & 206.0 & 94.5 & 0.97 & 0.06 &   &   & 1.61 &   \\  $q_{3}$ & 4.88 & 340.8 & 302.9 & 2.8 & 0.02 &   &   & 0.94 &   \\  $q_{4}$ & 4.82 & 325.6 & 268.0 & 1.81 & 0.02 &   & 0.87 & 0.48 & 0.41 \\  $q_{5}$ & 4.73 & 215.7 & 232.0 & 0.75 & 0.01 &   &   &   &   \\  $q_{6}$ & 4.88 & 413.7 & 58.1 & 0.92 & 0.02 &   &   & 0.64 & 0.77 \\  $q_{7}$ & 4.97 & 365.6 & 306.4 & 2.63 & 0.04 &   &   &   & 0.71 \\  $q_{8}$ & 4.9 & 304.7 & 223.0 & 0.9 & 0.02 &   &   &   & 0.72 \\  $q_{9}$ & 4.99 & 338.1 & 175.7 & 0.79 & 0.02 &   &   &   &   \\  $q_{10}$ & 4.83 & 338.7 & 268.3 & 1.93 & 0.01 &   &   & 0.64 & 0.61 \\  $q_{11}$ & 4.97 & 344.9 & 285.1 & 13.05 & 0.01 &   &   &   & 0.68 \\  $q_{12}$ & 4.94 & 286.7 & 194.8 & 1.24 & 0.02 &   & 0.45 &   &   \\  $q_{13}$ & 5.01 & 282.5 & 108.9 & 1.23 & 0.03 &   &   & 0.58 &   \\  $q_{14}$ & 4.9 & 306.1 & 114.5 & 1.56 & 0.02 & 0.45 & 0.47 &   &   \\  $q_{15}$ & 4.95 & 263.3 & 51.8 & 1.12 & 0.03 &   & 0.95 &   &   \\  $q_{16}$ & 4.96 & 138.4 & 14.1 & 1.48 & 0.03 & 0.48 & 3.57 &   &   \\  $q_{17}$ & 4.82 & 358.7 & 386.7 & 1.38 & 0.02 &   & 0.86 &   &   \\  $q_{18}$ & 4.79 & 279.6 & 100.6 & 9.17 & 0.02 &   &   &   & 0.51 \\  $q_{19}$ & 4.75 & 324.4 & 88.1 & 8.96 & 0.07 &   &   &   &   \\  $q_{20}$ & 4.86 & 162.8 & 104.2 & 1.32 & 0.07 &   & 3.34 & 2.03 &   \\  $q_{21}$ & 4.97 & 194.5 & 123.2 & 0.73 & 0.02 &   &   & 0.51 & 1.68 \\  $q_{22}$ & 5.04 & 231.1 & 187.9 & 1.46 & 0.06 &   &   &   & 1.28 \\  $q_{23}$ & 4.84 & 211.2 & 240.3 & 1.24 & 0.05 &   &   &   &   \\  $q_{24}$ & 5.1 & 233.0 & 178.1 & 20.19 & 0.02 &   & 0.74 & 1.29 &   \\  $q_{25}$ & 4.95 & 269.0 & 253.3 & 1.53 & 0.02 &   &   & 0.51 &   \\  $q_{26}$ & 4.85 & 180.3 & 2.8 & 0.82 & 0.07 &   &   & 1.2 &   \\  $q_{27}$ & 4.75 & 165.7 & 164.2 & 5.85 & 0.02 &   &   & 3.31 &   \\  $q_{28}$ & 4.99 & 140.3 & 89.9 & 3.51 & 0.03 &   & 0.91 & 1.3 & 1.26 \\  $q_{29}$ & 4.63 & 350.3 & 330.4 & 1.96 & 0.02 &   &   &   &   \\  $q_{30}$ & 4.73 & 216.2 & 60.5 & 0.66 & 0.02 &   &   & 0.7 & 0.57 \\  $q_{31}$ & 4.8 & 237.4 & 83.4 & 0.7 & 0.02 &   &   &   & 0.53 \\  $q_{32}$ & 4.91 & 250.5 & 230.6 & 0.84 & 0.02 &   & 0.54 &   &   \\  $q_{33}$ & 4.97 & 195.3 & 121.1 & 0.9 & 0.02 &   & 0.8 &   &   \\  $q_{34}$ & 4.68 & 308.5 & 146.3 & 2.93 & 0.01 &   & 1.5 &   &   \\  $q_{35}$ & 4.91 & 194.8 & 119.1 & 1.29 & 0.02 &   & 0.92 &   &   \\  $q_{36}$ & 4.81 & 158.5 & 245.4 & 1.49 & 0.01 &   & 0.5 &   &   \\  $q_{37}$ & 4.84 & 263.2 & 89.4 & 0.97 & 0.12 &   &   &   & 0.86 \\  $q_{38}$ & 4.78 & 283.1 & 154.6 & 1.82 & 0.03 &   &   &   &   \\  $q_{39}$ & 4.92 & 222.6 & 95.6 & 2.52 & 0.03 &   &   & 1.59 &   \\  $q_{40}$ & 4.87 & 224.4 & 24.4 & 3.96 & 0.02 &   &   & 1.14 & 1.07 \\  $q_{41}$ & 4.94 & 200.8 & 286.6 & 4.71 & 0.02 &   & 0.57 &   &   \\  $q_{42}$ & 5.06 & 161.2 & 113.1 & 0.98 & 0.02 &   &   & 0.41 & 1.21 \\  $q_{43}$ & 4.83 & 150.2 & 86.5 & 3.2 & 0.26 &   &   &   & 3.02 \\  $q_{44}$ & 4.95 & 319.1 & 23.8 & 2.27 & 0.03 &   &   &   & 0.8 \\  $q_{45}$ & 5.0 & 307.4 & 27.0 & 4.0 & 0.14 &   &   &   &   \\  $q_{46}$ & 4.85 & 196.1 & 153.3 & 1.22 & 0.04 &   &   & 1.55 & 3.05 \\  $q_{47}$ & 4.77 & 156.0 & 168.8 & 0.63 & 0.02 &   &   &   &   \\  $q_{48}$ & 4.84 & 276.5 & 70.4 & 1.54 & 0.02 &   &   & 0.95 & 1.45 \\  $q_{49}$ & 4.7 & 264.3 & 53.8 & 0.79 & 0.03 &   &   &   &   \\  $q_{50}$ & 4.78 & 337.6 & 44.0 & 1.32 & 0.04 &   &   & 1.69 & 1.07 
\end{tabular}
\end{ruledtabular}
\label{T1}
\end{table*}

\begin{table*}[ht]
\caption{\label{tab:brisbane2} Specifications of IBMQ Brisbane. }
\begin{ruledtabular}
\begin{tabular}{lccccccccc}
qubit & Frequency (GHz) & T1 ($\mu$s) & T2 ($\mu$s) & Readout Error (\%) & $\sqrt(X)$ (\%) & ECR $\uparrow$ (\%) & ECR $\downarrow$ (\%) & ECR $\leftarrow$ (\%) & ECR $\rightarrow (\%)$  \\   $q_{51}$ & 4.93 & 206.0 & 310.2 & 1.29 & 0.01 &   &   &   &   \\  $q_{52}$ & 4.92 & 202.1 & 191.2 & 1.0 & 0.05 & 6.84 & 0.6 &   &   \\  $q_{53}$ & 5.0 & 271.8 & 275.8 & 1.51 & 0.03 &   & 0.55 &   &   \\  $q_{54}$ & 4.97 & 196.4 & 125.0 & 2.02 & 0.03 & 1.24 & 0.92 &   &   \\  $q_{55}$ & 4.84 & 306.0 & 100.3 & 1.06 & 0.02 & 1.75 & 1.12 &   &   \\  $q_{56}$ & 4.81 & 205.0 & 161.2 & 1.1 & 0.02 &   &   &   & 0.61 \\  $q_{57}$ & 4.93 & 209.9 & 178.2 & 1.06 & 0.02 &   &   &   & 0.5 \\  $q_{58}$ & 4.89 & 161.8 & 65.6 & 21.25 & 0.02 &   & 0.57 &   & 1.06 \\  $q_{59}$ & 4.97 & 222.7 & 52.1 & 3.79 & 0.03 &   &   &   & 0.65 \\  $q_{60}$ & 4.77 & 285.4 & 127.5 & 3.39 & 0.05 &   &   &   & 1.28 \\  $q_{61}$ & 4.79 & 301.9 & 157.9 & 22.3 & 0.05 &   &   &   &   \\  $q_{62}$ & 4.94 & 132.7 & 163.2 & 1.63 & 0.02 &   & 0.86 & 1.41 & 1.11 \\  $q_{63}$ & 5.04 & 272.2 & 190.7 & 1.44 & 0.02 &   &   &   & 0.54 \\  $q_{64}$ & 4.82 & 284.2 & 151.9 & 0.85 & 0.02 &   &   &   &   \\  $q_{65}$ & 4.96 & 229.0 & 112.5 & 1.16 & 0.02 &   &   & 0.59 & 0.6 \\  $q_{66}$ & 4.89 & 192.5 & 140.7 & 0.97 & 0.02 &   &   &   &   \\  $q_{67}$ & 5.11 & 224.1 & 154.9 & 4.91 & 0.02 &   &   & 0.93 & 0.96 \\  $q_{68}$ & 4.74 & 224.8 & 31.6 & 2.99 & 0.03 &   &   &   &   \\  $q_{69}$ & 5.12 & 152.2 & 154.0 & 1.05 & 0.03 &   &   & 0.69 & 0.59 \\  $q_{70}$ & 4.9 & 345.3 & 161.9 & 7.07 & 0.02 &   &   &   &   \\  $q_{71}$ & 4.79 & 234.3 & 285.7 & 2.05 & 0.01 &   &   &   &   \\  $q_{72}$ & 5.06 & 202.8 & 96.8 & 4.17 & 0.03 &   &   &   &   \\  $q_{73}$ & 4.98 & 165.9 & 257.7 & 1.23 & 0.02 & 0.58 &   &   &   \\  $q_{74}$ & 5.03 & 256.6 & 110.1 & 1.07 & 0.04 & 1.09 & 1.18 &   &   \\  $q_{75}$ & 5.01 & 219.0 & 69.3 & 0.74 & 0.02 &   & 0.98 &   &   \\  $q_{76}$ & 4.84 & 300.1 & 65.3 & 1.13 & 0.01 &   &   & 0.62 &   \\  $q_{77}$ & 5.04 & 250.3 & 80.0 & 5.82 & 0.05 & 0.82 &   & 0.62 & 3.72 \\  $q_{78}$ & 4.64 & 167.6 & 262.2 & 1.21 & 0.02 &   &   &   &   \\  $q_{79}$ & 4.86 & 180.0 & 52.9 & 0.98 & 0.01 &   &   & 1.15 & 0.56 \\  $q_{80}$ & 5.03 & 192.1 & 55.2 & 5.48 & 2.51 &   &   &   & 0.6 \\  $q_{81}$ & 4.93 & 307.8 & 249.2 & 4.06 & 0.03 & 0.75 &   &   & 0.93 \\  $q_{82}$ & 4.89 & 84.3 & 115.7 & 1.54 & 0.1 &   &   &   & 1.14 \\  $q_{83}$ & 4.79 & 118.8 & 80.9 & 1.06 & 0.02 &   & 0.46 &   &   \\  $q_{84}$ & 4.68 & 249.4 & 142.5 & 0.86 & 0.01 &   &   & 0.61 &   \\  $q_{85}$ & 5.1 & 250.8 & 223.4 & 1.08 & 0.02 & 0.44 &   & 0.63 & 0.57 \\  $q_{86}$ & 4.9 & 139.7 & 84.6 & 4.14 & 0.04 &   &   &   & 1.64 \\  $q_{87}$ & 4.99 & 119.3 & 131.6 & 10.61 & 1.59 &   &   &   & 3.0 \\  $q_{88}$ & 5.1 & 135.8 & 85.7 & 0.61 & 0.03 &   &   &   & 0.86 \\  $q_{89}$ & 4.96 & 216.8 & 161.3 & 1.02 & 0.04 &   &   &   &   \\  $q_{90}$ & 4.91 & 230.0 & 84.0 & 1.07 & 0.05 &   &   &   &   \\  $q_{91}$ & 4.91 & 146.2 & 178.7 & 3.05 & 0.04 & 0.55 &   &   &   \\  $q_{92}$ & 4.91 & 120.1 & 137.4 & 0.55 & 0.02 &   & 0.57 &   &   \\  $q_{93}$ & 4.96 & 112.6 & 68.1 & 3.67 & 0.31 & 1.6 & 10.41 &   &   \\  $q_{94}$ & 4.98 & 146.3 & 151.7 & 1.1 & 0.02 & 1.14 &   &   & 0.61 \\  $q_{95}$ & 4.84 & 225.3 & 119.0 & 1.1 & 0.03 &   &   &   & 0.69 \\  $q_{96}$ & 4.75 & 300.2 & 78.6 & 4.91 & 0.03 &   &   &   &   \\  $q_{97}$ & 4.82 & 210.4 & 134.7 & 0.81 & 0.02 &   &   & 0.59 & 0.75 \\  $q_{98}$ & 4.94 & 115.5 & 25.1 & 1.83 & 0.11 & 1.28 &   &   &   \\  $q_{99}$ & 5.06 & 179.4 & 182.1 & 0.67 & 0.03 &   &   & 0.6 &   \\  $q_{100}$ & 4.91 & 149.5 & 225.3 & 0.92 & 0.02 &   & 0.86 & 3.38 &   
\end{tabular}
\end{ruledtabular}
\label{T1}
\end{table*}

\begin{table*}[ht]
\caption{\label{tab:brisbane3} Specifications of IBMQ Brisbane. }
\begin{ruledtabular}
\begin{tabular}{lccccccccc}
qubit & Frequency (GHz) & T1 ($\mu$s) & T2 ($\mu$s) & Readout Error (\%) & $\sqrt(X)$ (\%) & ECR $\uparrow$ (\%) & ECR $\downarrow$ (\%) & ECR $\leftarrow$ (\%) & ECR $\rightarrow (\%)$  \\    $q_{101}$ & 5.07 & 181.4 & 224.5 & 2.55 & 0.02 &   &   & 1.07 & 0.63 \\  $q_{102}$ & 4.81 & 213.7 & 160.3 & 4.16 & 0.02 &   &   &   & 0.87 \\  $q_{103}$ & 4.93 & 144.8 & 86.4 & 0.72 & 0.03 &   &   &   &   \\  $q_{104}$ & 4.77 & 196.3 & 176.0 & 1.29 & 0.02 &   &   & 0.71 &   \\  $q_{105}$ & 4.98 & 172.6 & 98.8 & 0.76 & 0.03 &   &   & 0.56 & 0.69 \\  $q_{106}$ & 4.88 & 219.4 & 186.2 & 1.68 & 0.03 &   &   &   &   \\  $q_{107}$ & 4.92 & 181.6 & 119.6 & 2.12 & 0.02 &   &   & 0.62 &   \\  $q_{108}$ & 5.06 & 163.6 & 160.9 & 2.15 & 0.03 &   & 0.73 & 0.68 &   \\  $q_{109}$ & 4.98 & 184.1 & 227.2 & 2.53 & 0.03 & 0.81 &   &   &   \\  $q_{110}$ & 4.83 & 202.2 & 276.2 & 1.19 & 0.03 &   & 0.86 &   &   \\  $q_{111}$ & 4.85 & 259.1 & 294.9 & 1.22 & 0.02 & 0.52 &   &   &   \\  $q_{112}$ & 5.0 & 176.1 & 67.2 & 0.47 & 0.02 &   & 0.79 &   &   \\  $q_{113}$ & 5.03 & 206.6 & 209.3 & 0.83 & 0.02 &   &   &   & 1.09 \\  $q_{114}$ & 4.82 & 185.2 & 61.0 & 13.86 & 0.02 & 0.61 &   &   & 0.98 \\  $q_{115}$ & 5.03 & 176.0 & 92.6 & 0.84 & 0.02 &   &   &   &   \\  $q_{116}$ & 4.91 & 205.8 & 191.0 & 0.77 & 0.01 &   &   & 0.57 & 0.56 \\  $q_{117}$ & 4.83 & 247.6 & 213.9 & 1.56 & 0.03 &   &   &   & 0.58 \\  $q_{118}$ & 4.73 & 241.4 & 116.4 & 0.76 & 0.02 &   &   &   & 0.99 \\  $q_{119}$ & 4.8 & 196.0 & 122.4 & 32.95 & 0.07 &   &   &   &   \\  $q_{120}$ & 4.84 & 184.5 & 154.6 & 17.81 & 0.03 &   &   & 1.58 &   \\  $q_{121}$ & 4.97 & 234.4 & 143.5 & 1.84 & 0.05 &   &   & 2.23 &   \\  $q_{122}$ & 4.94 & 224.1 & 130.2 & 2.21 & 0.05 & 0.75 &   & 1.36 & 0.84 \\  $q_{123}$ & 5.03 & 145.8 & 126.9 & 1.64 & 0.02 &   &   &   &   \\  $q_{124}$ & 4.95 & 214.2 & 194.8 & 0.82 & 0.03 &   &   & 0.59 &   \\  $q_{125}$ & 4.86 & 258.3 & 223.5 & 1.05 & 0.01 &   &   & 0.59 & 0.58
\end{tabular}
\end{ruledtabular}
\label{T1}
\end{table*}

\bibliography{ref}

\end{document}